\documentclass{article}

\usepackage[utf8]{inputenc}
\usepackage{amsmath,natbib}
\usepackage{hyperref}
\usepackage{url}
\usepackage{graphicx}
\usepackage{booktabs}
\usepackage{lipsum}
\usepackage{authblk}

\usepackage{xcolor}

\title{Validity in Music Information Research Experiments}
\author[1]{Bob L. T. Sturm}
\author[2]{Arthur Flexer}
\affil[1]{Division of Speech, Music and Hearing, KTH Stockholm, Sweden, \href{mailto:bobs@kth.se}{bobs@kth.se}}
\affil[2]{Institute of Computational Perception, Johannes Kepler University Linz, Austria, \href{mailto:arthur.flexer@jku.at}{arthur.flexer@jku.at}}

\date{}

\begin{document}
\maketitle
%%%%%%%%%%%%%%%%%%%%%%%%%%%%%%%%%%%%%%%%%%%%%%%%%%%%%%%%%%%%%%%%%%%%%%%%%%%%%%%%
% Abstract
%%%%%%%%%%%%%%%%%%%%%%%%%%%%%%%%%%%%%%%%%%%%%%%%%%%%%%%%%%%%%%%%%%%%%%%%%%%%%%%%
\vspace{-0.2in}
\begin{abstract}
Validity is the truth of an inference made from evidence,
such as data collected in an experiment, and is
central to working scientifically.
Given the maturity of the domain of music information research (MIR),
validity in our opinion should be discussed and considered 
much more than it has been so far.
Considering validity in one's work can improve its scientific and engineering value.
Puzzling MIR phenomena like adversarial attacks 
and performance glass ceilings become less mysterious through the lens of validity.
In this article, we review the subject of validity in general,
considering the four major types of validity from a key reference: \cite{shadish:etal:2002}.
We ground our discussion of these types with a prototypical MIR experiment: 
music classification using machine learning. 
Through this MIR experimentalists can be guided
to make valid inferences from data collected from their experiments.

\end{abstract}

%%%%%%%%%%%%%%%%%%%%%%%%%%%%%%%%%%%%%%%%%%%%%%%%%%%%%%%%%%%%%%%%%%%%%%%%%%%%%%%%
% Main Content Start
%%%%%%%%%%%%%%%%%%%%%%%%%%%%%%%%%%%%%%%%%%%%%%%%%%%%%%%%%%%%%%%%%%%%%%%%%%%%%%%%
\vspace{-0.2in}
\section{Introduction}

The multi-disciplinary field of Music Information Research (MIR) is focused on 
making music and information about music 
accessible to a variety of users.
This ranges from systems for search and retrieval, e.g., Shazam \citep{Wang2003} or TunePal \citep{Duggan2011a},
to recommendation, e.g., Spotify or Pandora,
and even to more creative applications like generation.
These systems consist of many computational components
working in domains between the acoustic and symbolic
where computability is possible \citep{Muller2015a}.
The effectiveness and reliability of these complex systems and their components
are of prime importance to the MIR researcher, 
not to mention other stakeholders.

The MIR researcher performs experiments to compare
approaches for modelling and retrieving music data.
A principal focus is on users,
but the cost of performing experiments with users is high,
and the replicability of such studies is difficult. 
This has motivated the use of computer-based experiments
where ``test collections'' serve as proxies for human users,
known as the {\em Cranfield Paradigm} \citep{Cleverdon:1991}:
the researcher predicts the usefulness of an MIR system 
by applying it to a representative sample of the problem domain
(e.g., queries and relevant documents).
While such an approach is inexpensive and replicable,
its relevance and reliability for MIR, and information retrieval in general, 
have been questioned \citep{Voorhees2001a,urbano:etal:2013}.

Under the Cranfield Paradigm,
state-of-the-art MIR systems perform exceptionally
well in reproducing the ground truth of some datasets, 
e.g., inferring rhythm, genre or emotion from audio data.
However, slight and irrelevant transformations of the audio
can suddenly render these systems ineffectual 
\citep{sturm:2014,kereliuk:etal:2015,Rodriguez-Algarra2016a,prinz2021end}.
In one case \citep{Rodriguez-Algarra2016a}, a ``genre recognition'' system relies on 
infrasonic signatures, seemingly originating 
from the data collection, but nonetheless imperceptible and
irrelevant for human listeners.
In another case  \citep{Sturm2015a}, a ``rhythm recognition'' system 
seemingly uses tempo to infer rhythm, a confusion
originating again from the data collection.

Related are discussions about ``glass ceilings'' 
of MIR systems \citep{Aucouturier2004,Pohle2008},
i.e., that the observed barrier to improving system performance
to perfect or human level is due to the psychophysical and cultural factors of music 
missing from computable features extracted from audio recordings \citep{wiggins:2009}.
MIR researchers striving for great improvement of those numbers
in established problems and datasets
might actually make no practical difference for human users \citep{urbano2012significant}.
\cite{flexer:grill:2016} identifies one
reason contributing to performance glass ceilings: 
the data used to train and test MIR systems
can arise from tasks that are not well-posed (e.g., music similarity), 
and thus render meaningless
the pursuit of better numbers using that data.
The lack of definition in research problems motivates
\cite{sturm2014kiki} to propose MIR researchers solve synthetic yet well-defined simplifications
of harder, ill-posed problems.

At the heart of an MIR experiment is the relationship between
conclusions drawn from its results and their {\em validity}, or ``truth value'' \citep{shadish:etal:2002}.
Ideally, an experiment will be carefully designed and implemented to answer
a well-defined hypothesis.\footnote{We encourage readers to review \cite{Chase2001} to see the remarkable lengths an experimentalist must go to test even a simple hypothesis.}
Its components -- units, treatments, design, observations, and settings --
should be operationalised (translated from theory into practice) 
to maximize quality and minimize cost (money, time, ethics).
This is the purview of the discipline {\em Design of Experiments}:
how can one get the best evidence for the least cost?
A look through the proceedings of MIR's premier conference ISMIR\footnote{\href{https://ismir.net/conferences/}{https://ismir.net/conferences/}} and related publications, however, 
reveals a general lack of awareness of validity in the MIR community.

The first reference explicitly introducing validity to MIR is \cite{Urbano2011b}, 
discussing it in relation to text information retrieval (IR) literature, 
and specifically advocating validity-based meta-evaluation of MIREX results.\footnote{‘Music Information Retrieval Evaluation eXchange’ (MIREX) is an annual evaluation campaign for MIR algorithms \citep{downie2006music}.} 
This is further developed in \cite{urbano:etal:2013}, 
embracing the validity typology of \cite{shadish:etal:2002}, and
bringing in notions of reliability and efficiency.
While the review of \cite{Schedl:etal:2013} 
does not explicitly refer to validity or even mentions the Cranfield Paradigm,
it makes clear how central computer experiments are to MIR,
and criticizes a general lack of consideration of users
in its conclusions.
\cite{Sturm:2013} asserts that just reporting figures of merit like accuracy is not sufficient to decide whether an MIR system is really recognizing ``genre'' in musical signals, or whether it relies on irrelevant confounding factors, later stating that these uncontrolled factors are a danger to the validity of conclusions drawn from such experiments \citep{sturm:2014,Sturm:2017}.
\cite{Sturm2014b} attempts to define music description (including the ``use case'') 
to motivate evaluating music description systems in ways that allow for valid and relevant conclusions.

More recently,
\cite{rodriguez2019characterising} propose and test a method for 
detecting the existence of confounding in MIR experiments,
thus addressing the lack of construct validity of conclusions drawn from them.
\cite{liem2020can} studies construct validity of high-level musical concepts like genre 
by confronting MIR systems with data different from what they have been trained on. 
\cite{flexer2021evaluation} criticizes the lack of external validity of experiments on general music similarity and demands focus on specific aspects of music similarity and definition of a specific use case. 
Probably the most recent attempt to comprehensively discuss validity and reliability in MIR is an unpublished  tutorial held at ISMIR 2018 \citep{urbano2018statistical}.

The above review of previous work on validity in MIR
shows it to be dispersed and fragmented across only relatively few publications.
Despite a small chorus of calls to address major methodological problems
of MIR experiments to improve validity in the discipline, e.g., \cite{Urbano2011b,Schedl:etal:2013,
sturm:2014,urbano2018statistical,liem2020can,flexer2021evaluation},
there has yet to be published a systematic and critical engagement
of what validity means in the context of MIR,
and how to consider it when designing, implementing and analyzing
experiments.

In this article, we review the four principal types of validity in \cite{shadish:etal:2002}, 
an authoritative resource about validity in causal inference and experimental science.
Other typologies exist, e.g., \cite{Lund2021q}, 
but we focus on that of \cite{shadish:etal:2002}
because it is an established point of reference,
and has already been introduced to MIR \citep{urbano:etal:2013}.
For each type, we discuss common threats,
relate it to the MIR discipline in general,
and then more concretely in terms of a typical MIR experiment,
presented in Sec. \ref{Sec:typical}.
We include a python jupyter notebook that runs 
the experiment and computes a variety of statistics and results discussed below.\footnote{\url{https://github.com/boblsturm/mirvaliditytutorial}\label{footnote:github}}
Section \ref{sec:experiment} reviews the components of the experiment,
on which the discussion of validity hinges.
Each section \ref{sec:statistical_validity}--\ref{sec:external_validity} reviews the four types of validity and presents actionable questions which can help MIR researchers to scrutinize the conclusions they draw from their experiments.

\section{A Typical MIR Experiment}\label{Sec:typical}
To ground our general discussion of validity,
we present a typical MIR experiment 
that exemplifies a considerable amount of MIR research:
audio classification using machine learning
and a benchmark dataset.
Consider the BALLROOM dataset \citep{Dixon2004},
created around 2004 from downloading excerpts of music CDs for
sale at a website focused on ballroom dancing.
BALLROOM consists of 698 audio recordings, each about 30 seconds long
and labeled with one of eight dance styles or music rhythms, 
e.g., ``Waltz''.
BALLROOM has appeared in dozens of studies \citep{Sturm:2017},
and most often in experiments measuring the amount of ground truth labels reproduced
by different combinations of classifiers and features.

We create a random 70/30 partition,
with 488/210 recordings comprising the training/testing dataset.
From each recording we compute 
the spectral flux onset strength envelope 
using a hop size of 1024 samples, or 46 ms \citep{McFee2015b}.
We compute the normalised autocorrelation of the envelope
and retain the portion relating to lags in $[0.23, 4.14]$ seconds.
Finally, we model the autocorrelation by a 12th-order autoregressive model,
which results in a feature of 13 dimensions describing the 30-second audio recording.
We normalise each dimension of the training features
to be zero mean and unit variance.
We normalise the testing data features using the same parameters 
as for normalising the training data features.

\begin{table}[t]
\footnotesize\centering
\begin{tabular}{c|c|c|c|c|}
 & Accuracy & Precision & Recall & f1-score \\ \hline
LDA & $0.714$ & $0.711$ & $0.711$ & $0.703$ \\ \hline
QDA & $0.719$ & $0.715$ & $0.723$ & $0.717$ \\ \hline
1NN & $0.662$ & $0.644$ & $0.635$ & $0.638$ \\ \hline
3NN & $0.681$ & $0.673$ & $0.651$ & $0.656$ \\ \hline
5NN & $0.719$ & $0.699$ & $0.687$ & $0.689$ \\ \hline
7NN & $0.695$ & $0.669$ & $0.656$ & $0.659$ \\ \hline
9NN & $0.700$ & $0.681$ & $0.664$ & $0.668$ \\ \hline
unif &$0.12\pm 0.02$ &$0.13\pm 0.03$ &$0.12\pm 0.02$ &$0.12\pm 0.02$\\ \hline
freq &$0.13\pm 0.02$ &$0.13\pm 0.03$ &$0.13\pm 0.02$ &$0.13\pm 0.02$\\ \hline
maj &$0.16$ &$0.02$ &$0.12$ &$0.03$\\ \hline
\end{tabular}
\caption{Accuracy, and macro-averaged precision, recall and f1-score observed for several models in a testing partition of BALLROOM \citep{Dixon2004}. The performance of two models randomly selecting labels (with standard deviation) are shown in the rows labeled: {\em unif} samples labels uniformly; {\em freq} samples labels according to training data label frequency. 
The last row {\em maj} shows the performance of a model choosing the label most frequent in the training data.}
\label{tab:modelsBALLROOM}
\end{table}

We model features using multivariate Gaussian distributions
(linear and quadratic discriminant analysis, LDA and QDA), or k-nearest neighbours (KNN).
We compare the labels inferred by a model to the ground truth 
and compute the accuracy, and macro-averaged precision, recall, 
and f1-score.\footnote{A macro-averaged figure of merit 
is the mean figure of merit measured across classes, 
regardless of the number of observations of each class.}
Table \ref{tab:modelsBALLROOM} shows the outcomes of this experiment,
which can be reproduced with 
the supplementary python jupyter notebook.\footnote{See footnote \ref{footnote:github}.}
The last rows show statistics of three other strategies:
{\em unif} samples labels according to a uniform prior and
{\em freq} samples labels according to training data label frequency;\footnote{The mean and standard deviation of these figures of merit can be found in theory, but here they are found empirically. See the accompanying code in footnote \ref{footnote:github}.}
{\em maj} selects the most frequent label in the training data.
These provide comparison points.

A criticism can be made here that the above experiment is not really representative
of MIR experiments today: datasets are now several orders of magnitude larger,
and deep learning has in many cases made obsolete the engineering of features.
However, we will press on with this example for four reasons.
First, the BALLROOM dataset is analyzed to such an extent that classification
experiments with it illustrate the four types of validity in clear and graspable ways.
Using a larger dataset can illustrate the same things, 
but would first require a thorough analysis of its composition and provenance.
Second, concerns about validity are not rendered moot 
by simply increasing data or shifting the machine learning approach.
Third, the experiment above is typical in its {\em design}: a dataset is collected and processed,
submitted to a variety of machine learning approaches, 
and measurements are made by applying the trained systems
to some test collection.
Finally, the BALLROOM dataset is unique in that 
an ``extended'' version of it was created  
a decade later from the same web resource \citep{Marchand2016a}. 
This allows us to meaningfully test the external validity of conclusions
about systems trained in BALLROOM, e.g., the concepts learned from
BALLROOM are relevant to ballroom dance music, and not just
ballroom dance music in 2004 (see Sec. 7.2).

\section{Components of an Experiment}\label{sec:experiment}
Before discussing the validity of conclusions drawn from the typical MIR experiment, 
we must identify its units, treatments, design, observations, and settings.
{\em Treatments} are the things applied to units
in order to cause an effect (or not in the case of a {\em control}),
{\em units} are the things that are treated,
and {\em observations} are what is measured on a unit.
The {\em design} specifies which treatment is applied to which unit,
and {\em settings} involve time, place, and condition.
To make this more concrete, consider a medical experiment
in which the effect of a treatment on blood pressure is being studied.
A number of people are sampled from a population, 
some of whom will receive the treatment
while the others receive a placebo (control).
The design describes which people get the treatment, 
and which do not.
The observation is the blood pressure of a person after one month.
The setting can include particulars of the population (rural or urban), 
place of treatment (hospital or home), and so on.
The experimentalist contrasts blood pressure observations
across groups to conclude if the treatment 
has an effect.

Returning to our typical MIR experiment, 
we wish to determine the effectiveness of
different machine learning (ML) algorithms 
in predicting the labels of a test recording dataset.
There are two possibilities here.
We can see the treatments as the ten algorithms,
and the units as the entire testing dataset 
(how does the dataset respond to each algorithm?),
or we can see the entire testing dataset as the one treatment,
and the units as the ten algorithms 
(how does each algorithm respond to the dataset?).
Since Table \ref{tab:modelsBALLROOM} reports figures of merit (observations) 
of each algorithm on the entire dataset, the latter interpretation
motivates conclusions about the effectiveness of particular algorithms.
In this case, the design is simple: 
each unit is given the same treatment.
The setting involves the partitioning of BALLROOM, the features extracted, random seeds, software libraries, and so on.

\vspace{-0.1in}
\section{Statistical Conclusion Validity}
\label{sec:statistical_validity}
Statistical conclusion validity is 
``the validity of inferences about covariation between two variables''
\citep{shadish:etal:2002}.
This includes concluding that a covariation exists, 
and perhaps its strength as well.
This is the level at which one is concerned with {\em statistical significance},
i.e., that an observed covariation between treatment and effect
is not likely to arise by chance. 
\cite{shadish:etal:2002} (p.\ 45) includes a table of nine different threats to statistical conclusion validity.
Four threats relevant to computer-based experiments are: violated assumptions about the statistics underlying the responses
(and the use of the wrong statistical test, a {\em type III error} \citep{Kimball1957});
a sample size too small to reliably detect covariation (lack of power);
the purposeful search for significant results
by trying multiple analyses and data selections (``p-hacking'', \cite{p_hacking});
and increased variance in observations due to the heterogeneity of units.

\subsection{MIR concerns about statistical conclusion validity}
 
{\em Are my results statistically significant?} Null hypothesis statistical testing (NHST) quantifies 
whether the observed effects of the treatments on the responses arise by mere chance,
as well as the direction of effect and its size.
This answers the question: are the results statistically significant?
Fundamentals about statistical testing in MIR are discussed by \cite{Flexer:2006},
for Artificial Intelligence in general by \cite{cohen1995empirical}, and for machine learning by \cite{Japkowicz2011}.
One must take care in selecting a statistical test to use;
each one makes strong assumptions that could be violated.
NHST is most straightforwardly applicable to completely randomized experimental designs \citep{Bailey2008}, 
thereby reducing the possibility of
structure in units and treatments interfering with the responses (which results in confounding, discussed in the next section).
Most MIR experiments cannot use complete randomisation because the target population from which samples come is unclear
(what is a random sample of ``sad'' music, with the term ``sad'' being quite ill-defined?),
and so the kinds of conclusions that can be made with NHST in MIR are limited.\footnote{Experimental designs that cannot be completely randomised are called {\em quasi-experimental designs}, 
which is another major topic of \cite{shadish:etal:2002}.}

{\em Is the observed statistical significance relevant for a user?} In MIR, even if one finds statistical significance, 
this may not generalise to a perceivable difference for 
actual users interacting with the ``improved'' MIR system.
As an example from MIR, a crowd-sourced user evaluation \citep{urbano2012significant} demonstrates that there is an upper bound of user satisfaction with music recommendation systems of about $80\%$, since this was the highest percentage of users agreeing that two systems ``are equally good.'' In addition, for the MIREX task of {\em Audio Music Similarity and Retrieval}  \cite{urbano2012significant} demonstrate that statistically significant differences between algorithms can be so small that they make no practical difference for users.

\subsection{Statistical conclusion validity in the typical MIR experiment}
Let us now consider the typical MIR experiment in Sec. \ref{Sec:typical}
and reason about what conclusions we can draw from it
that have statistical conclusion validity.
Table \ref{tab:modelsBALLROOM} clearly shows
that each response of machine learning (ML) to the dataset
is greater than the random approaches 
{\em unif}, {\em freq} and {\em maj}.
How likely is it that any of the responses of ML
is due to chance, i.e., that any of the ML approaches
is actually no better than one of the random approaches?
Since we have the empirical distributions for 
{\em unif} and {\em freq}, we can estimate
the probability of either of them resulting in, e.g., 
a macro-average recall at least as large as $0.6$: $p<e^{-200}$.\footnote{
This is the probability of observing a macro-averaged recall at least as big as 0.6 
from the models selecting labels randomly.
Specifically, it is the probability of sampling 
a number from a random variable distributed Gaussian with 
mean $0.125$ and standard deviation $0.02$
in the domain $[0.6, \infty)$. See the accompanying jupyter notebook. 
Also note that a macro-averaged recall of 0.6 is a pessimistic choice. 
Using the higher numbers in Table \ref{tab:modelsBALLROOM} results in even lower probability.}
Hence, a valid statistical conclusion is that 
we observe a significant covariation between the use of ML
with these particular features and the responses measured 
on a specific partition of BALLROOM.
An equivalent conclusion is that each response of these trained ML systems treated by this partition of BALLROOM are inconsistent with choosing randomly.

Whereas the conclusions above relate to the use of ML,
one might consider statistical conclusions relating to
the type of ML, i.e., Gaussian modelling (LDA and QDA) vs. 
nearest neighbour modelling (KNN), or LDA vs.\ QDA.
One may want to conclude that
Gaussian modelling is better than nearest neighbour modelling
with these features on BALLROOM,
or even more precisely,
QDA performs the best with these features on BALLROOM.
This involves looking at the differences between responses,
called {\em contrasts}.
While there is not much reason to suspect that
the particular 70/30 partition of BALLROOM used
is unusually beneficial to one kind of ML model over the other,
what we do not know from our experiment 
is the distribution of any response due to ML, 
and so of any difference of responses due to ML,
related to the random partitioning of BALLROOM.
We have two measurements of Gaussian models,
and five of nearest neighbour models, 
but we cannot simply compute and compare statistics of these
without involving auxiliary information not present in the experiment,
e.g., how the decision boundaries of an ML model 
vary according to variation in the training data.

In fact, if we conclude from Table \ref{tab:modelsBALLROOM} that Gaussian modelling
performs better than nearest neighbour modelling with these features 
on 70/30 partitions of BALLROOM, we would be wrong.
This is a ``type I error'',
which is concluding there to be a significant difference when in fact there is none.
When we perform this experiment 1000 times with random 70/30 partitions 
we observe that the difference between the best response of a Gaussian model
and the best response of a nearest neighbour model is distributed
Gaussian, and that the probability of observing zero difference or less is $p>0.41$ for any of the figures of merit (see the accompanying jupyter notebook). 
That is, the more statistically valid conclusion is that 
three out of five times randomly partitioning BALLROOM
the best Gaussian model
performs better than the best nearest neighbor model.
With this new data we see that the results in Table \ref{tab:modelsBALLROOM} 
do not merit a valid statistical conclusion
that one of the modelling approaches is significantly 
better or worse than any other in BALLROOM using these features
for any significance level $\alpha<0.4$.
On the other hand, if we were to conclude that QDA 
performs better than nearest neighbour modelling with these features 
on 70/30 random partitions of BALLROOM,
we would be correct -- but only for macro-averaged recall.
Performing 1000 repetitions of our experiment shows that the distribution
of the difference between the response from QDA
and the best response of all others is distributed
such that the probability of observing a negative difference 
is $p < 0.046$ only for recall (see the accompanying jupyter notebook).
In other words, less than one out of 20 times do we see
QDA perform worse than the other models in terms of recall.

In summary, the most general statistical conclusion 
we can make from Table \ref{tab:modelsBALLROOM}
is that the responses we observe from ML are highly inconsistent with 
the responses of models choosing randomly.
Each ML model knows {\em something} about BALLROOM
linking the features computed from a music recording
with its label.
Because we do not know the amount of variation in any response
due to partitioning in Table \ref{tab:modelsBALLROOM}, 
we cannot make any valid statistical conclusion
about which type of ML model is the best, 
or which particular realisation is the best,
for this particular dataset.
In order to go further, we had to run the experiment
multiple times to obtain distributions of the contrasts.
Even then, however, we cannot say anything about
the {\em cause} of significant differences yet.
This is where the notion of internal validity becomes relevant.

\section{Internal Validity}
\label{sec:internal_validity}
Internal validity is ``the validity of inferences about whether the 
observed covariation between two variables is causal''
\citep{shadish:etal:2002}.
While statistical conclusion validity is concerned only with
the strength of covariation between treatment and responses,
internal validity is focused on the {\em cause} of
a particular response to the treatment.
\cite{shadish:etal:2002} (p. 55) includes a table of nine different threats to causal conclusions.
Several of these involve {\em confounding}, which is the confusion of the treatment with other factors
arising from poor operationalisation in an experiment.

As a concrete example, consider an experiment measuring 
the effects of two different medicines on lowering blood pressure,
but where one medicine is given to young patients
and the other is given to elderly patients.
This experimental design confounds the two medicines
and patient age, and so the effects caused by the two factors
cannot be disambiguated.
Any conclusion from this experiment
about the effects of the medicines
lacks internal validity.

\subsection{MIR concerns about internal validity}

{\em Does my data collection introduce confounds?} One's methodology for collecting music data might unintentionally introduce structure.
For instance, \cite{Sturm:2017} discusses how the BALLROOM dataset was assembled by downloading excerpts of music CDs sold at a website selling music for ballroom dance competitions.
Ballroom dance competitions are regulated by organisations, e.g., World DanceSport Federation (WDSF),\footnote{\href{https://www.worlddancesport.org/}{https://www.worlddancesport.org/}} to ensure uniformity of events for competitors around the world.
These organisations set strict requirements of tempo of each dance such that high skill is required of the dancers.
Hence, the labels of the BALLROOM dataset can mean: 
1) the rhythm of the music; 2) the type of dance;
3) the strict tempo requirements of the dance.

{\em Does my data partitioning introduce confounds?} Dataset partitioning can also introduce confounds, e.g., ``bleeding ground truth.''
An example is to first segment recordings into short (e.g., 40ms) time frames 
and then partition these frames into training and testing sets, thus 
spreading highly correlated features across these sets.
In the context of audio-based genre classification, the presence of songs from the same artists or albums in both training and test data has been shown to artificially inflate performance \citep{pampalk2005improvements,flexer2010effects}.
\cite{flexer2007closer} shows that audio-based genre classification using very direct representations of spectral content 
degrade more when employing artist/album filters than classification based on more abstract kind of features like rhythmic content (fluctuation patterns). 
This insight that problems of data partitioning can affect MIR systems in quite different ways and hence change performance rankings has been confirmed in another meta-study \citep{sturm2014jnmr}.

Research on explainable and interpretable MIR might help to identify confounds, see e.g., \citep{mishra2018understanding}, but latest results have shown that many popular explainers struggle when being probed with deliberate confounding transformations \citep{Praher2021Veracity,hoedt2022plausibility}. 
A general framework for detecting and 
characterising effects of confounding in MIR experiments through 
interventions is proposed by \cite{sturm:2014}
and further refined in \cite{rodriguez2019characterising}. 
More work has yet to be done in the domain of
causal inference in MIR, but is difficult
because its core concepts, like ``similarity'',
need to be operationalised first. 
Otherwise disentangling variables potentially influencing measurements is not possible.

\subsection{Internal validity in the typical MIR experiment}\label{sec:IVMIR}
Of interest is {\em what} it is in our trained ML models of Sec. \ref{Sec:typical}
causing their response to be inconsistent with random selection.
Knowing how Gaussian models used in LDA and QDA are built --
mean and covariance parameters are estimated from training data --
an internally valid conclusion is that these models
work well in BALLROOM because likelihood distributions 
estimated from the training data also fit the testing data well.
Similarly, knowing how nearest neighbour classifiers work,
an internally valid conclusion is that in BALLROOM the neighbourhoods
of the labeled features in the training data include to a large
extent testing data features with the same labels.
Another internally valid conclusion is  
that the high performances of these 
ML models in BALLROOM are caused by the features
together with the expressivity of the models
capturing information related to the labels in BALLROOM.
Our ML models have learned {\em something} about BALLROOM,
which causes their performance to be significantly better
than random selection.

With reference to the aims of MIR research,
we want to conclude something more specific, e.g., our ML models have learned to recognize the rhythms in BALLROOM.
This is certainly one explanation consistent with
the performance of our systems, but is it the only one?
The internal validity of this conclusion relies on a key assumption:
inferring the correct labels of BALLROOM 
can {\em only} be the result of learning to discriminate between
and identify the rhythms in BALLROOM.
In other words, we must assume that there is no other way to
accurately infer labels in BALLROOM than by perceiving rhythm.

It is not difficult to imagine other ways
to infer the labels in BALLROOM by 
considering the multitude of variables present
in the music recordings.
One can hear instrumentation
unique to each class, e.g.,
many recordings labeled ``Tango''
feature strings, piano and accordion.
Many recordings labeled ``Waltz''
and ``Viennese Waltz'' feature strings.
Many recordings labeled ``Quickstep'' and ``Jive''
feature brass, piano, vocals, and drums.
So perceiving instrumentation is one way to 
infer labels in BALLROOM with more success
than random selection.
We can reject such an explanation knowing that
the time-domain nature of the features
we are using are not sensitive to timbre,
and so cannot clearly express the sound characteristic of an instrument.
Then, are there time-domain variables other than rhythm
that are closely associated with the labels in BALLROOM?

One possible variable is tempo.
If tempo is correlated with rhythm in BALLROOM
then tempo estimation is another way
a ML model can reproduce the ground truth labels.
From listening to BALLROOM recordings  
labeled ``Waltz'', ``Quickstep'' or ``Cha cha cha'',
one can hear they feature 
different rhythms and different tempi,
but discriminating between them can be done
on the basis of tempo alone.
Tempo and rhythm are related musical characteristics,
but they are not one and the same thing \citep{Sethares2007x}.

\begin{figure}[t]
 \centerline{
 \includegraphics[width=0.7\columnwidth]{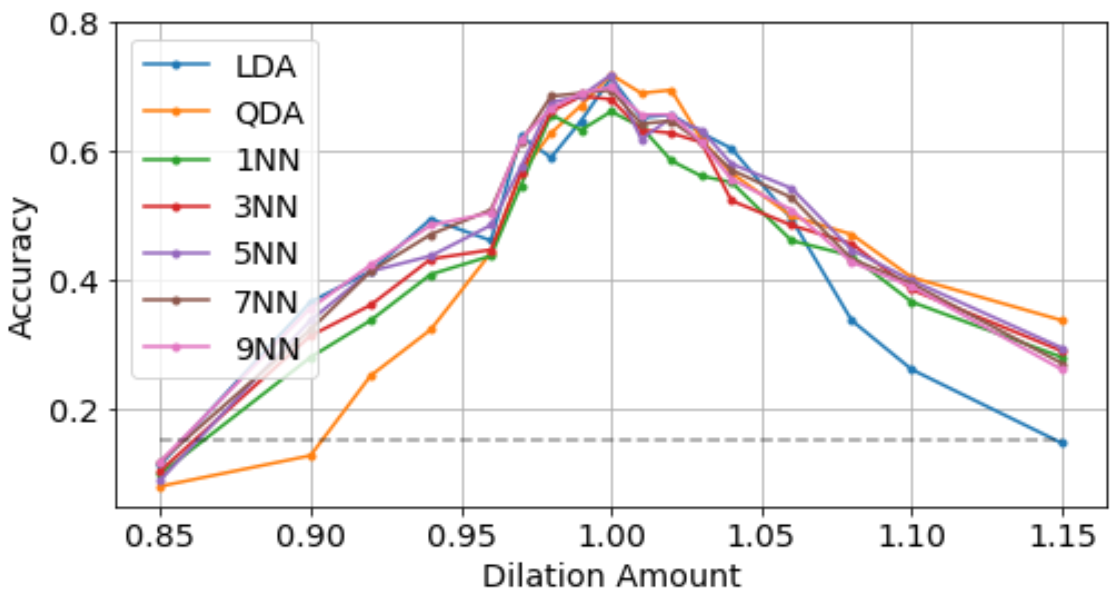}}
 \caption{Accuracies of the ML models in Table \ref{tab:modelsBALLROOM}
 with test recordings undergoing pitch-preserving time dilation.
 Horizontal dashed line in grey is the mean accuracy of the best random system plus twice
 its standard deviation.}
 \label{fig:AccTempo}
\end{figure}

Let us perform an experiment to test the sensitivity of
our trained ML models to tempo.
We perform an intervention where we alter all test recordings
by some amount of pitch-preserving time dilation,
and then measure the responses of the models
to these new ``treatments''.
Dilating a recording by an amount 1.1 increases its duration
by 10\% -- or equivalently, makes the music in the recording
have a tempo that is 10\% slower.
Dilating a recording by an amount 0.9 decreases its duration
by 10\% -- or equivalently, makes the music in the recording
have a tempo that is 10\% faster.
Figure \ref{fig:AccTempo} shows how the
accuracy of all ML models from Table \ref{tab:modelsBALLROOM}
covaries with the amount of dilation.
Our intervention has clearly revealed the extent to which 
the ML models rely on the tempi in the test data --
which is no surprise given the background information of BALLROOM in the previous subsection.

The validity of the conclusion that the responses of
our ML models are caused by their
recognition of the rhythms in BALLROOM
relies on too strong of an assumption about BALLROOM,
i.e., that rhythm recognition is the only way to infer labels in BALLROOM.
The experimental design does not account 
for the structure present in the dataset;
we do not control for other
ways of inferring the labels of BALLROOM,
which are guaranteed to exist by its very construction.
From Table \ref{tab:modelsBALLROOM}
and our experimental design, 
we thus cannot be any more specific in our
causal inference than this: the responses of our 
ML models are caused by their
having learned {\em something} about BALLROOM.
At best, they are relying on at least tempo and rhythm.
This then calls into question how comparing
predictions with ground truth in BALLROOM
relates to the ability we might actually want to measure,
that is the recognition of rhythm.
This is where the notion of construct validity
becomes relevant.

\section{Construct Validity}
\label{sec:construct_validity}
Construct validity is ``the validity of inferences about the 
higher order constructs that represent sampling particulars''
\citep{shadish:etal:2002}.
This involves the relationship between what is meant to be inferred by the experimentalist from an experiment
and what is actually measured, i.e., 
the {\em operationalisation} of the experimentalist's intention.
For instance, directly measuring the blood pressure
of a person involves an invasive procedure inserting a measuring device in their veins.
Blood pressure can be measured less invasively but indirectly by externally applying known pressure 
to a vein and listening for when blood flow ceases.
Knowledge about the incompressibility of liquids in closed systems
makes the measurement of pressure in the balloon a relevant measure of blood pressure.
\cite{shadish:etal:2002} (p. 73) includes a table of fourteen different threats to construct validity, 
but several of these are irrelevant to computer-based experiments.
The main threat is a questionable relationship between what is being measured and what is intended to be measured. 
Selecting a measure by convenience but not relevance, 
sampling from convenient populations,
and a lack of definition of what is intended to be measured,
are threats to construct validity.
Construct validity involves more than just how something is measured;
it also involves what is measured and in what settings.

\subsection{MIR concerns about construct validity}

{\em How is classification accuracy, or any figure of merit, in a labeled music dataset related to {\em X}?} Two examples in MIR are the use of ``genre'' classification accuracy
as an indirect measure of music similarity \citep{Pohle2008}, 
or IR user satisfaction (see, e.g., \citep{Schedl:etal:2013} for a discussion).
The relationship between these is very tenuous, 
especially so considering that accuracy itself is an
unreliable measure of whether or not a system has learned
anything relevant to music \citep{Sturm:2013,sturm:2014}.
A key reference in this respect 
is that of \cite{Pfungst1911} describing a series of experiments
in trying to reliably measure the arithmetic acumen 
of a horse that was only able to tap out answers.
Counting the number of correct answers tapped out by the horse,
no matter how many questions are asked, 
is irrelevant without considering how each question is posed (the setting).
The key to Pfungst discovering the cause
of the horse's apparent arithmetic acumen
involved changing the setting:
the questions remained the same,
and accuracy of correct response was measured,
but how the questions were posed was changed
in order to control for different factors of the experiment.
The same is true for MIR.

{\em What is the ``use case'' of the system to be tested?}
To counter threats to construct validity 
the MIR experimentalist must operationalise as much as possible 
the use case of the system to be built and tested.
An attempt to do so for music description is in \cite{Sturm2014b}, which emphasises the need to 
define success criteria.
The experimentalist must determine how their
method of measurement relates to the success criteria,
e.g., relating classification accuracy in BALLROOM
to the satisfaction of a specific user.

{\em How can we test the construct validity of a conclusion?}
One possibility is to assess the outcomes of different experiments which are supposed to measure the same higher order constructs. 
An example in MIR is to study correlations of different genre classifiers when given identical inputs \citep{liem2020can}. 
Low correlations between classifiers point to problems of construct validity.
A related topic is that of adversarial examples,
which casts doubt on the conclusion that 
the high accuracy of an MIR system in some dataset 
reflects its ``perception'' of the music in the waveform.
Adversarial examples have first been described in image analysis \citep{Szegedy2014IntruigingProperties}, where imperceptible perturbations of input data significantly degraded classification accuracy. 

For music genre classification systems, irrelevant audio filtering transformations of music signals are used in \cite{sturm:2014} to both deflate and inflate classification accuracy to be no better than chance level or perfect $100\%$. The irrelevance is ascertained with listening tests and real human subjects, 
with the transformation being audible but not changing the clear impression of a certain musical genre.
The same problematic behavior has been documented concerning music emotion classification \citep{sturm:2014}.
For deep music embedding a test based on imperceptible audio transformations has been proposed \citep{kim2019nearby}, essentially verifying distance consistency both in the input audio
space and corresponding latent deep space.

Following these so-called untargeted attacks which try to change a prediction to an arbitrary target, targeted attacks aiming at changing predictions to specific classes have been explored. A targeted attack on genre recognition has been reported \citep{kereliuk:etal:2015}, where magnitude spectral frames computed from audio are treated as images and attacked using approaches from image object recognition. For music instrument classification a targeted attack allowing to add perturbations directly to audio waveforms instead of spectrograms has also been presented \citep{prinz2021end}. Quite similar to the results by \citep{kereliuk:etal:2015}, the attacks were able to reduce the accuracy close to a random baseline and produce misclassifications to any desired instrument. Signal perturbations were almost imperceptible apart from some high-frequency deviations. The authors also artificially boosted playcounts via an attack on a real-world music recommender, thereby demonstrating that such attacks can be a security issue in MIR. Follow-up work presented lines of defence against such malicious attacks \citep{Hoedt2022DefenceSoundpark}.

\subsection{Construct validity in the typical MIR experiment}
When it comes to our typical MIR experiment in Sec. \ref{Sec:typical}, 
we are interested in making construct inferences
around the latent ability of rhythm recognition we are supposedly measuring.
For instance, one construct inference is that
our features measure relevant aspects of rhythm in recorded music.
In some sense, by their definition from basic signal processing
components, our features come from temporal aspects that
are certainly relevant to rhythm.
Our features are also reliant on acoustic information,
and in particular there being high-contrast differences
in onsets captured by spectral flux -- hence limiting their
relationship to rhythms played by 
particular kinds of instruments
with sharp attacks.
However, we have seen 
above that the features are also indicative of tempo,
which is not rhythm \citep{Sethares2007x},
and that tempo is another path an ML algorithm
can use to get to the rhythm label.
Hence we are left to question the relationship of our features
to the concept we are trying to operationalise, i.e., rhythm.

The construction of the BALLROOM dataset,
intended to reflect different ballroom dance rhythms,
is closely related to the validity of construct inferences
derived from its use. 
How do the ``Waltz'' excerpts exemplify the ``Waltz'' rhythm?
Is there one ``Waltz'' rhythm?
In BALLROOM, there are actually two different labels for waltz:
``Waltz'' and ``Viennese Waltz.''
The distinction between them is based in part on tempo,
according to the World Sport Dance Federation,
a Viennese waltz is to be performed at a tempo between
174-180 BPM \cite{WSDF2014}.

Having a system label any partition of the BALLROOM dataset
provides no reliable measure of a system's ability to recognise rhythm without changing the setting to control for other factors.
It is not as simple as choosing a different feature, measure, 
cross-validation method, or using a particular statistical test.
One must change the experiment itself such that
{\em rhythm recognition} is what is actually being measured.
This means that BALLROOM can still be useful
to measuring the rhythm recognition of a ML system.
Indeed, in the previous section
we used it to disprove the causal claim
that the good performance of the ML systems of Table \ref{tab:modelsBALLROOM} 
is caused by their ability to recognize rhythm.
Might performance in BALLROOM also be an indication
of performance in other datasets focused on rhythm?
This is where the notion of external validity becomes relevant.

\section{External Validity}
\label{sec:external_validity}
External validity is ``the validity of inferences about the 
extent to which a causal relationship holds over variations in 
experimental units, settings, treatment variables and measurement variables'' \citep{shadish:etal:2002}.
More generally, external validity is the truth of   
a generalised causal inference drawn from an experiment. 
An example is inferring that medicine found to lower blood pressure in patients living in Germany 
will also lower blood pressure in people living in Mexico  --
a conclusion that can lack validity due to differences 
in diet, living and working conditions, and so on.
Another example is that increasing the dose of the 
medicine will cause blood pressure to lower further
in the studied population.
If a causal inference we draw from an experiment lacks internal validity, then generalising that inference to include variations not tested will not have external validity.
\cite{shadish:etal:2002} (p. 87) includes a table of five different threats to external validity,
which are in addition to the threats to internal validity.
The main threat is that variation of the components of the experiment
might destroy the causal inference that holds in the experiment. 
For instance, a medication may work for the type of illness tested, but that type of illness may not be generalisable to other closely related illnesses.

\subsection{MIR concerns about external validity}

{\em Does my model generalize to out-of-sample data?} The standard approach in evaluating MIR classification systems is to use separate train and test sets in cross-validation experiments to obtain seemingly unbiased estimates of performance. However, if such MIR systems are exposed to independent out-of-sample data often severe loss of performance is observed. One example are experiments on genre recognition where accuracy results do not hold when evaluated across different collections that are not part of the training sets \citep{bogdanov2016cross,bogdanov2019acousticbrainz}. 
The results do not generalize to supposedly identical genre labels in different collections, which reflects a lack of external validity. Genre labels like e.g.\ 'Rock' will be used differently by different annotators working on these collections 
-- which is a threat to construct validity. 
Another example are how different audio encodings affect subsequent computation of descriptors and classification results \citep{urbano2014effect}, or how in general differences in software implementations diminish replicability \citep{mcfee2018open}.

{\em Do different raters agree on a ground truth?} Human perception of music is highly subjective resulting in possible low inter-rater agreement. Therefore only a certain amount of agreement can be expected if several human subjects are asked to rate the same song pairs according to their perceived similarity, depending on a number of subjective factors \citep{Schedl:etal:2013,flexer:grill:2016} like personal taste, listening history, familiarity with the music, current mood, etc. Concerning annotation of music, \cite{Seyerlehner:etal:2010} shows that the performance of humans classifying songs into 19 genres ranges from modest $26\%$ to $71\%$. Audio-based grounding of everyday musical terms shows  the same problematic results \citep{Aucouturier:2009}. It has even been argued \citep{wiggins:2009} that no such thing as an immovable `ground' exists in the context of music, because music itself is subjective, highly context-dependent and dynamic.

The lack of inter-rater agreement presents a problem of external validity because inferences from the experiment do not generalize from users or annotators in the experiment to the intended target population of arbitrary users/annotators. 
It is also a problem of reliability, since different groups of users or annotators with their differing subjective opinions will impede repeatability of experimental results. 
This lack of inter-rater agreement presents an upper bound for MIR approaches, since it is not meaningful to have computational models going beyond the level of human agreement. Such upper bounds have been reported \citep{Jones:etal:2007,Schedl:etal:2013,flexer:grill:2016} for the MIREX tasks of `Audio Music Similarity and Retrieval' (AMS) and `Music Structural Segmentation' (MSS). For AMS the upper bound has already been reached in 2009, while for MSS the upper bound is within reach for at least some genres of music. Comparable results exist concerning  music structure analysis \citep{Nieto:etal:2014} and chord estimation \citep{Ni:etal:2013,Koops:2019}. 

{\em Do raters agree with themselves at different points in time?} Going beyond the question of whether different annotators agree on a ground truth one can also access what the level of agreement within one person is when faced with identical annotation tasks at different points in time. A high intra-rater agreement would help to overcome the problem of upper bounds in MIR systems since it would make personalization of models meaningful, i.e.\
to have separate models for individual persons. However, at least for the task of general music similarity it has been shown that intra-rater agreement is only slightly higher than inter-rater agreement \citep{flexer2021evaluation}, with the absolute level also depending on music material and mood of raters at test time. An approach to personalize chord labels for individual annotators via deep learning was more successful \citep{koops2020automatic}.

Returning to the impact of irrelevant transformations \citep{sturm:2014,kereliuk:etal:2015,Rodriguez-Algarra2016a}
and the existence of adversarial examples \citep{prinz2021end}, 
one can ask, does my model generalize to these kinds of attacks?
More constructively, one can seek ways to lessen the impact
of these attacks, thus possibly increasing the generalization
of the models \citep{Hoedt2022DefenceSoundpark}.

\subsection{External validity in the typical MIR experiment}
Considering the typical MIR experiment in Sec. \ref{Sec:typical},
we cannot validly conclude that any of our
models is recognizing rhythm in general
because we do not know if they are 
recognizing rhythm in BALLROOM.
Our dilation intervention experiment
in Sec. \ref{sec:IVMIR}
reveals that all of the models lose their
supposed ability to recognize rhythm in BALLROOM,
so there is no reason to infer they will
recognize rhythm elsewhere.
One causal conclusion we might generalise
is that all our models perform well in BALLROOM
because they have learned something about BALLROOM --- 
a curated set of recordings downloaded from a specific website in 2004.
Might our models have also learned something about other recordings from that same website but collected many years later?

The extended BALLROOM dataset (X-BALLROOM) \citep{Marchand2016a} consists of 3,484 audio recordings
in the same eight dance styles or music rhythms as BALLROOM,
but downloaded from the same website over a decade later.
This gives us a chance to test our hypothesis.
The figures of merit measured from our models trained in BALLROOM
but applied to all of X-BALLROOM
are shown in Table \ref{tab:modelsExtendedBALLROOM}.
We still see significant covariation between response and
the use of ML with our features.
By and large, whatever concepts our ML models
have learned about BALLROOM carry over
to X-BALLROOM -- but we still do not know whether or not those concepts
have to do with rhythm.

\begin{table}[t]
\footnotesize\centering
\begin{tabular}{c|c|c|c|c|}
 & Accuracy & Precision & Recall & f1-score \\ \hline
LDA & $0.659$ & $0.647$ & $0.643$ & $0.643$ \\ \hline
QDA & $0.682$ & $0.678$ & $0.672$ & $0.673$ \\ \hline
1NN & $0.622$ & $0.616$ & $0.602$ & $0.604$ \\ \hline
3NN & $0.636$ & $0.629$ & $0.610$ & $0.613$ \\ \hline
5NN & $0.644$ & $0.643$ & $0.617$ & $0.619$ \\ \hline
7NN & $0.647$ & $0.646$ & $0.619$ & $0.621$ \\ \hline
9NN & $0.645$ & $0.643$ & $0.615$ & $0.618$ \\ \hline
unif &$0.12\pm 0.01$ &$0.13\pm 0.01$ &$0.12\pm 0.01$ &$0.12\pm 0.01$\\ \hline
freq &$0.13\pm 0.01$ &$0.13\pm 0.01$ &$0.12\pm 0.01$ &$0.12\pm 0.01$\\ \hline
maj &$0.13$ &$0.02$ &$0.12$ &$0.03$\\ \hline
\end{tabular}
\caption{As in Table \ref{tab:modelsBALLROOM}, models trained in BALLROOM and tested in all of X-BALLROOM \citep{Marchand2016a}.}
\label{tab:modelsExtendedBALLROOM}
\end{table}

We can take this opportunity to test
the generalizability of a conclusion about
Gaussian modelling outperforming
nearest neighbor modelling from
Table \ref{tab:modelsBALLROOM}.
Training and testing the same ML models
with a 70/30 random partition of
X-BALLROOM produces the results in 
Table \ref{tab:modelsExtendedBALLROOM2}.
We still see QDA performing the best,
but the nearest neighbor models  
show large performance increases.
The cause of this change in performance
should be investigated, and related to the
confounding known to exist in BALLROOM.
Does the confounding also exist in X-BALLROOM,
suggested by the results in Table \ref{tab:modelsExtendedBALLROOM}?
This shows how the observations made in 
the typical MIR experiment represent the
beginning of avenues for exploration,
sources of hypotheses, and not the final result.

\begin{table}[t]
\footnotesize\centering
\begin{tabular}{c|c|c|c|c|}
Mod & Accuracy & Precision & Recall & f1-score \\ \hline
LDA & $0.726$ & $0.721$ & $0.722$ & $0.717$ \\ \hline
QDA & $0.763$ & $0.764$ & $0.773$ & $0.758$ \\ \hline
1NN & $0.702$ & $0.691$ & $0.693$ & $0.690$ \\ \hline
3NN & $0.741$ & $0.731$ & $0.727$ & $0.727$ \\ \hline
5NN & $0.750$ & $0.744$ & $0.736$ & $0.738$ \\ \hline
7NN & $0.753$ & $0.746$ & $0.736$ & $0.738$ \\ \hline
9NN & $0.751$ & $0.746$ & $0.734$ & $0.737$ \\ \hline
unif &$0.12\pm 0.01$ &$0.13\pm 0.01$ &$0.12\pm 0.01$ &$0.12\pm 0.01$\\ \hline
freq &$0.13\pm 0.01$ &$0.13\pm 0.01$ &$0.13\pm 0.01$ &$0.12\pm 0.01$\\ \hline
maj &$0.13$ &$0.02$ &$0.12$ &$0.03$\\ \hline
\end{tabular}
\caption{As in Table \ref{tab:modelsBALLROOM}, but models trained and tested in X-BALLROOM \citep{Marchand2016a}.}
\label{tab:modelsExtendedBALLROOM2}
\end{table}

\section{Conclusion}
This article provides a review of 
the notion of validity based on the typology 
given in \cite{shadish:etal:2002}.
It brings together the few sources in MIR that 
mention anything to do with validity, 
and several sources that do not but are related.
This article does not aim to prescribe 
how to design and perform experiments
such that valid conclusions can be drawn from them.
Instead it aims to bring within the realm
of MIR what validity means, why it is important,
and how it can be threatened.

In MIR the predominant experimental methodology is 
given by the Cranfield paradigm: train a model on a partition of a dataset 
and count the number of correct answers on another partition.
This kind of experiment is inexpensive to do with the data conveniently at hand,
and provides numbers that can be 
compared in ways that convince peer reviewers that
progress has been accomplished \citep{Hand2006}.
Despite various appeals \citep{Peeters2012,Schedl:etal:2013} and beseechings \citep{urbano2012significant,urbano:etal:2013,Sturm:2013,sturm:2014,flexer:grill:2016,flexer2021evaluation},
such an experimental approach is still standard in the field 
and its serious flaws are ignored. 
Any inference from this experiment
that is more general than ``the system has learned something 
about this dataset'' lacks internal, construct and external validity.
This does not mean that all such inferences are false,
just that they cannot follow from the experiment
as designed and implemented.
Reproducing the ground truth of a dataset represents a beginning and must be followed by a search for the causes of the observed behavior.
One must resist the urge to conclude that a system must be doing whatever is hoped for.

\cite{shadish:etal:2002} provides an
established starting point for MIR,
but there exist other types of validity.
For instance, \cite{Lund2021q} revises the typology of \cite{shadish:etal:2002}
to address ambiguities between causes and treatments,
to better define aspects of settings, and to establish
a hierarchical ordering of five types of validity:
statistical conclusion, causal, construct, 
generalization and theoretical.
An important distinction in this typology 
from that of \cite{shadish:etal:2002} is its 
emphasis on a major aim of basic research: to contribute theory.
Other kinds of validity include ecological, convergent, and criterion \citep{Urbano2011b};
but these still deal with the kind of
conclusion one is drawing from evidence
collected in some way.

As a final note, a frustration when encountering \cite{shadish:etal:2002} 
as an engineer is that of its 623 pages there are only five pages with at least one equation on them. Instead, \cite{shadish:etal:2002} describe experiments and how each type of validity manifests in the conclusions drawn, with specific threats to the reasoning of those conclusions.
Experiments, not to mention experimentalists, are such complex assemblages that 
expressing them in formal ways that appear to permit the computation
of numbers that relate to each type of validity would probably
have very limited applicability, and then only be understood by a limited audience.
The language of validity is {\em reason}, 
and we hope this manuscript will inspire
MIR researchers to think creatively
about the phenomena they observe to discover
their causes.

\section*{Acknowledgments}
We thank Juli\'an Urbano and Hugo Maruri-Aguilar for helpful discussions
during the drafting of this manuscript, as well as the constructive criticisms
of reviewers from the Transactions of the Society for Music Information Retrieval.
The contribution of Sturm is supported by a project that has received funding from 
the European Research Council (ERC) under the European Union’s Horizon 2020 
research and innovation programme (Grant agreement No. 864189 MUSAiC: 
Music at the Frontiers of Artificial Creativity and Criticism).
The contribution of Flexer is supported by funding from the Austrian Science Fund (FWF, project number P 31988).
For the purpose of open access, the authors have applied a CC BY public copyright license to any author accepted manuscript version arising from this submission.

\bibliography{ARXIVtemplate}

\begin{thebibliography}{65}
\providecommand{\natexlab}[1]{#1}
\providecommand{\url}[1]{\texttt{#1}}
\expandafter\ifx\csname urlstyle\endcsname\relax
  \providecommand{\doi}[1]{doi: #1}\else
  \providecommand{\doi}{doi: \begingroup \urlstyle{rm}\Url}\fi

\bibitem[Aucouturier(2009)]{Aucouturier:2009}
J.-J. Aucouturier.
\newblock Sounds like teen spirit: Computational insights into the grounding of
  everyday musical terms.
\newblock \emph{Language, evolution and the brain}, pages 35--64, 2009.

\bibitem[Aucouturier and Pachet(2004)]{Aucouturier2004}
J.-J. Aucouturier and F.~Pachet.
\newblock Improving timbre similarity: How high is the sky?
\newblock \emph{J. Neg. Results Speech Audio Sci.}, 1\penalty0 (1):\penalty0
  1--13, 2004.

\bibitem[Bailey(2008)]{Bailey2008}
R.~A. Bailey.
\newblock \emph{Design of comparative experiments}.
\newblock Cambridge University Press, 2008.

\bibitem[Bogdanov et~al.(2016)Bogdanov, Porter, Boyer, Serra,
  et~al.]{bogdanov2016cross}
D.~Bogdanov, A.~Porter, H.~Boyer, X.~Serra, et~al.
\newblock Cross-collection evaluation for music classification tasks.
\newblock In \emph{Proceedings of the 17th International Society for Music
  Information Retrieval Conference}, pages 379 -- 385, 2016.

\bibitem[Bogdanov et~al.(2019)Bogdanov, Porter, Schreiber, Urbano, and
  Oramas]{bogdanov2019acousticbrainz}
D.~Bogdanov, A.~Porter, H.~Schreiber, J.~Urbano, and S.~Oramas.
\newblock The acousticbrainz genre dataset: Multi-source, multi-level,
  multi-label, and large-scale.
\newblock In \emph{Proceedings of the 20th Conference of the International
  Society for Music Information Retrieval}, 2019.

\bibitem[Chase(2001)]{Chase2001}
A.~Chase.
\newblock Music discriminations by carp ``{C}yprinus carpio''.
\newblock \emph{Animal Learning \& Behavior}, 29\penalty0 (4):\penalty0
  336--353, 2001.

\bibitem[Cleverdon(1991)]{Cleverdon:1991}
C.~W. Cleverdon.
\newblock The significance of the cranfield tests on index languages.
\newblock In \emph{Proceedings of the 14th annual international ACM SIGIR
  conference on Research and development in information retrieval}, pages
  3--12, 1991.

\bibitem[Cohen(1995)]{cohen1995empirical}
P.~R. Cohen.
\newblock \emph{Empirical methods for artificial intelligence}, volume 139.
\newblock MIT press Cambridge, MA, 1995.

\bibitem[Dixon et~al.(2004)Dixon, Gouyon, and Widmer]{Dixon2004}
S.~Dixon, F.~Gouyon, and G.~Widmer.
\newblock Towards characterisation of music via rhythmic patterns.
\newblock In \emph{Proceedings of the 5th Conference of the International
  Society for Music Information Retrieval}, pages 509--517, 2004.

\bibitem[Downie(2006)]{downie2006music}
J.~S. Downie.
\newblock The music information retrieval evaluation exchange (mirex).
\newblock \emph{D-Lib Magazine}, 12\penalty0 (12), 2006.

\bibitem[Duggan and O'Shea(2011)]{Duggan2011a}
B.~Duggan and B.~O'Shea.
\newblock Tunepal: searching a digital library of traditional music scores.
\newblock \emph{OCLC Systems \& Services}, 27:\penalty0 284--297, 2011.

\bibitem[Federation(2014)]{WSDF2014}
W.~S.~D. Federation.
\newblock \emph{WSDF Competition Rules}.
\newblock Bucharest, Romania, 2014.

\bibitem[Flexer(2006)]{Flexer:2006}
A.~Flexer.
\newblock Statistical evaluation of music information retrieval experiments.
\newblock \emph{Journal of New Music Research}, 35\penalty0 (2):\penalty0
  113--120, 2006.

\bibitem[Flexer(2007)]{flexer2007closer}
A.~Flexer.
\newblock A closer look on artist filters for musical genre classification.
\newblock In \emph{Proceedings the 8th International Conference on Music
  Information Retrieval}, pages 341--344, 2007.

\bibitem[Flexer and Grill(2016)]{flexer:grill:2016}
A.~Flexer and T.~Grill.
\newblock The problem of limited inter-rater agreement in modelling music
  similarity.
\newblock \emph{Journal of new music research}, 45\penalty0 (3):\penalty0
  239--251, 2016.

\bibitem[Flexer and Schnitzer(2010)]{flexer2010effects}
A.~Flexer and D.~Schnitzer.
\newblock Effects of album and artist filters in audio similarity computed for
  very large music databases.
\newblock \emph{Computer Music Journal}, 34\penalty0 (3):\penalty0 20--28,
  2010.

\bibitem[Flexer et~al.(2021)Flexer, Lallai, and
  Ra{\v{s}}l]{flexer2021evaluation}
A.~Flexer, T.~Lallai, and K.~Ra{\v{s}}l.
\newblock On evaluation of inter-and intra-rater agreement in music
  recommendation.
\newblock \emph{Transactions of the International Society for Music Information
  Retrieval}, 4\penalty0 (1):\penalty0 182--194, 2021.

\bibitem[Hand(2006)]{Hand2006}
D.~J. Hand.
\newblock Classifier technology and the illusion of progress.
\newblock \emph{Statistical Science}, 21\penalty0 (1):\penalty0 1--15, 2006.

\bibitem[Head et~al.(2015)Head, Holman, Lanfear, Kahn, and Jennions]{p_hacking}
M.~L. Head, L.~Holman, R.~Lanfear, A.~T. Kahn, and M.~D. Jennions.
\newblock The extent and consequences of p-hacking in science.
\newblock \emph{PLOS Biology}, 13\penalty0 (3):\penalty0 1--15, 2015.

\bibitem[Hoedt et~al.(2022{\natexlab{a}})Hoedt, Flexer, and
  Widmer]{Hoedt2022DefenceSoundpark}
K.~Hoedt, A.~Flexer, and G.~Widmer.
\newblock {Defending a Music Recommender Against Hubness-Based Adversarial
  Attacks}.
\newblock In \emph{Proceedings of the 19th Sound and Music Computing
  Conference}, pages 385--390, 2022{\natexlab{a}}.

\bibitem[Hoedt et~al.(2022{\natexlab{b}})Hoedt, Praher, Flexer, and
  Widmer]{hoedt2022plausibility}
K.~Hoedt, V.~Praher, A.~Flexer, and G.~Widmer.
\newblock Constructing adversarial examples to investigate the plausibility of
  explanations in deep audio and image classifiers.
\newblock \emph{Neural Computing and Applications}, 2022{\natexlab{b}}.

\bibitem[Japkowicz and Shah(2011)]{Japkowicz2011}
N.~Japkowicz and M.~Shah.
\newblock \emph{Evaluating Learning Algorithms: A Classification Perspective}.
\newblock Cambridge University Press, New York, NY, USA, 2011.

\bibitem[Jones et~al.(2007)Jones, Downie, and Ehmann]{Jones:etal:2007}
M.~C. Jones, J.~S. Downie, and A.~F. Ehmann.
\newblock Human similarity judgments: Implications for the design of formal
  evaluations.
\newblock In \emph{Proceedings of the 8th Conference of the International
  Society for Music Information Retrieval}, pages 539--542, 2007.

\bibitem[Kereliuk et~al.(2015)Kereliuk, Sturm, and Larsen]{kereliuk:etal:2015}
C.~Kereliuk, B.~L. Sturm, and J.~Larsen.
\newblock Deep learning and music adversaries.
\newblock \emph{IEEE Transactions on Multimedia}, 17\penalty0 (11):\penalty0
  2059--2071, 2015.

\bibitem[Kim et~al.(2019)Kim, Urbano, Liem, and Hanjalic]{kim2019nearby}
J.~Kim, J.~Urbano, C.~Liem, and A.~Hanjalic.
\newblock Are nearby neighbors relatives? {T}esting deep music embeddings.
\newblock \emph{Frontiers in Applied Mathematics and Statistics}, page~53,
  2019.

\bibitem[Kimball(1957)]{Kimball1957}
A.~W. Kimball.
\newblock Errors of the third kind in statistical consulting.
\newblock \emph{J. American Statistical Assoc.}, 52\penalty0 (278):\penalty0
  133--142, June 1957.

\bibitem[Koops et~al.(2019)Koops, De~Haas, Burgoyne, Bransen, Kent-Muller, and
  Volk]{Koops:2019}
H.~V. Koops, W.~B. De~Haas, J.~A. Burgoyne, J.~Bransen, A.~Kent-Muller, and
  A.~Volk.
\newblock Annotator subjectivity in harmony annotations of popular music.
\newblock \emph{Journal of New Music Research}, 48\penalty0 (3):\penalty0
  232--252, 2019.

\bibitem[Koops et~al.(2020)Koops, de~Haas, Bransen, and
  Volk]{koops2020automatic}
H.~V. Koops, W.~B. de~Haas, J.~Bransen, and A.~Volk.
\newblock Automatic chord label personalization through deep learning of shared
  harmonic interval profiles.
\newblock \emph{Neural Computing and Applications}, 32\penalty0 (4):\penalty0
  929--939, 2020.

\bibitem[Liem and Mostert(2020)]{liem2020can}
C.~C. Liem and C.~Mostert.
\newblock Can't trust the feeling? how open data reveals unexpected behavior of
  high-level music descriptors.
\newblock In \emph{Proceedings of the 21st International Society for Music
  Information Retrieval Conference}, pages 240--247, 2020.

\bibitem[Lund(2021)]{Lund2021q}
T.~Lund.
\newblock A revision of the campbellian validity system.
\newblock \emph{Scandinavian J. Educational Research}, 65\penalty0
  (3):\penalty0 523--535, 2021.

\bibitem[Marchand and Peeters(2016)]{Marchand2016a}
U.~Marchand and G.~Peeters.
\newblock Scale and shift invariant time/frequency representation using
  auditory statistics: Application to rhythm description.
\newblock In \emph{IEEE International Workshop on Machine Learning for Signal
  Processing}, 2016.

\bibitem[McFee et~al.(2015)McFee, Raffel, Liang, Ellis, McVicar, Battenberg,
  and Nieto]{McFee2015b}
B.~McFee, C.~Raffel, D.~Liang, D.~P.~W. Ellis, M.~McVicar, E.~Battenberg, and
  O.~Nieto.
\newblock librosa: Audio and music signal analysis in python.
\newblock In \emph{Proceedings Python in Science Conference}, pages 18--25,
  2015.

\bibitem[McFee et~al.(2018)McFee, Kim, Cartwright, Salamon, Bittner, and
  Bello]{mcfee2018open}
B.~McFee, J.~W. Kim, M.~Cartwright, J.~Salamon, R.~M. Bittner, and J.~P. Bello.
\newblock Open-source practices for music signal processing research:
  Recommendations for transparent, sustainable, and reproducible audio
  research.
\newblock \emph{IEEE Signal Processing Magazine}, 36\penalty0 (1):\penalty0
  128--137, 2018.

\bibitem[Mishra et~al.(2018)Mishra, Sturm, and Dixon]{mishra2018understanding}
S.~Mishra, B.~L. Sturm, and S.~Dixon.
\newblock Understanding a deep machine listening model through feature
  inversion.
\newblock In \emph{Proceedings of the 19th International Society for Music
  Information Retrieval Conference}, pages 755--762, 2018.

\bibitem[M\"uller(2015)]{Muller2015a}
M.~M\"uller.
\newblock \emph{Fundamentals of Music Processing: Audio, Analysis, Algorithms,
  Applications}.
\newblock Springer, 2015.

\bibitem[Ni et~al.(2013)Ni, McVicar, Santos-Rodriguez, and
  De~Bie]{Ni:etal:2013}
Y.~Ni, M.~McVicar, R.~Santos-Rodriguez, and T.~De~Bie.
\newblock Understanding effects of subjectivity in measuring chord estimation
  accuracy.
\newblock \emph{IEEE Transactions on Audio, Speech, and Language Processing},
  21\penalty0 (12):\penalty0 2607--2615, 2013.

\bibitem[Nieto et~al.(2014)Nieto, Farbood, Jehan, and Bello]{Nieto:etal:2014}
O.~Nieto, M.~M. Farbood, T.~Jehan, and J.~P. Bello.
\newblock Perceptual analysis of the f-measure for evaluating section
  boundaries in music.
\newblock In \emph{Proceedings of the 15th International Society for Music
  Information Retrieval Conference}, pages 265--270, 2014.

\bibitem[Pampalk et~al.(2005)Pampalk, Flexer, Widmer,
  et~al.]{pampalk2005improvements}
E.~Pampalk, A.~Flexer, G.~Widmer, et~al.
\newblock Improvements of audio-based music similarity and genre classificaton.
\newblock In \emph{Proceedings of the 6th Conference of the International
  Society for Music Information Retrieval}, pages 634--637, 2005.

\bibitem[Peeters et~al.(2012)Peeters, Urbano, and Jones]{Peeters2012}
G.~Peeters, J.~Urbano, and G.~J.~F. Jones.
\newblock Notes from the {ISMIR} 2012 late-breaking session on evaluation in
  music information retrieval.
\newblock In \emph{Proceedings of the 13th Conference of the International
  Society for Music Information Retrieval}, 2012.

\bibitem[Pfungst(1911)]{Pfungst1911}
O.~Pfungst.
\newblock \emph{Clever Hans (The horse of Mr. Von Osten): A contribution to
  experimental animal and human psychology}.
\newblock Henry Holt, New York, 1911.

\bibitem[Pohle et~al.(2008)Pohle, Pampalk, and Widmer]{Pohle2008}
T.~Pohle, E.~Pampalk, and G.~Widmer.
\newblock Evaluation of frequently used audio features for classification of
  music into perceptual categories.
\newblock In \emph{International Workshop Content-Based Multimedia Indexing},
  2008.

\bibitem[Praher et~al.(2021)Praher, Prinz, Flexer, and
  Widmer]{Praher2021Veracity}
V.~Praher, K.~Prinz, A.~Flexer, and G.~Widmer.
\newblock {On the Veracity of Local, Model-agnostic Explanations in Audio
  Classification: Targeted Investigations with Adversarial Examples}.
\newblock In \emph{Proceedings of the 22nd International Society for Music
  Information Retrieval Conference}, pages 531--538, 2021.

\bibitem[Prinz et~al.(2021)Prinz, Flexer, and Widmer]{prinz2021end}
K.~Prinz, A.~Flexer, and G.~Widmer.
\newblock On end-to-end white-box adversarial attacks in music information
  retrieval.
\newblock \emph{Transactions of the International Society for Music Information
  Retrieval}, 4\penalty0 (1):\penalty0 93--104, 2021.

\bibitem[Rodr\'iguez-Algarra et~al.(2016)Rodr\'iguez-Algarra, Sturm, and
  Maruri-Aguilar]{Rodriguez-Algarra2016a}
F.~Rodr\'iguez-Algarra, B.~L. Sturm, and H.~Maruri-Aguilar.
\newblock Analysing scattering-based music content analysis systems: {W}here's
  the music?
\newblock In \emph{Proceedings of the 17th Conference of the International
  Society for Music Information Retrieval}, pages 344--350, 2016.

\bibitem[Rodr{\'\i}guez-Algarra et~al.(2019)Rodr{\'\i}guez-Algarra, Sturm, and
  Dixon]{rodriguez2019characterising}
F.~Rodr{\'\i}guez-Algarra, B.~Sturm, and S.~Dixon.
\newblock Characterising confounding effects in music classification
  experiments through interventions.
\newblock \emph{Transactions of the International Society for Music Information
  Retrieval}, pages 52--66, 2019.

\bibitem[Schedl et~al.(2013)Schedl, Flexer, and Urbano]{Schedl:etal:2013}
M.~Schedl, A.~Flexer, and J.~Urbano.
\newblock The neglected user in music information retrieval research.
\newblock \emph{Journal of Intelligent Information Systems}, 41\penalty0
  (3):\penalty0 523--539, 2013.

\bibitem[Sethares(2007)]{Sethares2007x}
W.~A. Sethares.
\newblock \emph{Rhythm and Transforms}.
\newblock Springer, 2007.

\bibitem[Seyerlehner et~al.(2010)Seyerlehner, Widmer, and
  Knees]{Seyerlehner:etal:2010}
K.~Seyerlehner, G.~Widmer, and P.~Knees.
\newblock A comparison of human, automatic and collaborative music genre
  classification and user centric evaluation of genre classification systems.
\newblock In \emph{International Workshop on Adaptive Multimedia Retrieval},
  pages 118--131. Springer, 2010.

\bibitem[Shadish et~al.(2002)Shadish, Cook, and Campbell]{shadish:etal:2002}
W.~R. Shadish, T.~D. Cook, and D.~T. Campbell.
\newblock \emph{Experimental and quasi-experimental designs for generalized
  causal inference}.
\newblock Boston: Houghton Mifflin, 2002.

\bibitem[Sturm(2013)]{Sturm:2013}
B.~L. Sturm.
\newblock Classification accuracy is not enough.
\newblock \emph{Journal of Intelligent Information Systems}, 41\penalty0
  (3):\penalty0 371--406, 2013.

\bibitem[Sturm(2014{\natexlab{a}})]{sturm2014jnmr}
B.~L. Sturm.
\newblock The state of the art ten years after a state of the art: Future
  research in music information retrieval.
\newblock \emph{Journal of New Music Research}, 43\penalty0 (2):\penalty0
  147--172, 2014{\natexlab{a}}.

\bibitem[Sturm(2014{\natexlab{b}})]{sturm:2014}
B.~L. Sturm.
\newblock A simple method to determine if a music information retrieval system
  is a ``horse''.
\newblock \emph{IEEE Transactions on Multimedia}, 16\penalty0 (6):\penalty0
  1636--1644, 2014{\natexlab{b}}.

\bibitem[Sturm(2017)]{Sturm:2017}
B.~L. Sturm.
\newblock The ``horse'' inside: seeking causes behind the behaviors of music
  content analysis systems.
\newblock \emph{Computers in Entertainment (CIE)}, 14\penalty0 (2):\penalty0
  1--32, 2017.

\bibitem[Sturm and Collins(2014)]{sturm2014kiki}
B.~L. Sturm and N.~Collins.
\newblock The kiki-bouba challenge: Algorithmic composition for content-based
  mir research \& development.
\newblock In \emph{Proceedings of the 15th Conference of the International
  Society for Music Information Retrieval}, pages 21--26, 2014.

\bibitem[Sturm et~al.(2014)Sturm, Bardeli, Langlois, and Emiya]{Sturm2014b}
B.~L. Sturm, R.~Bardeli, T.~Langlois, and V.~Emiya.
\newblock Formalizing the problem of music description.
\newblock In \emph{Proceedings of the 15th Conference of the International
  Society for Music Information Retrieval}, pages 89--94, 2014.

\bibitem[Sturm et~al.(2015)Sturm, Kereliuk, and Larsen]{Sturm2015a}
B.~L. Sturm, C.~Kereliuk, and J.~Larsen.
\newblock ?`{El Caballo Viejo}? {L}atin genre recognition with deep learning
  and spectral periodicity.
\newblock In \emph{Proceedings of the International Conference on Mathematics
  and Computation in Music}, pages 335--346, 2015.

\bibitem[Szegedy et~al.(2014)Szegedy, Zaremba, Sutskever, Bruna, Erhan,
  Goodfellow, and Fergus]{Szegedy2014IntruigingProperties}
C.~Szegedy, W.~Zaremba, I.~Sutskever, J.~Bruna, D.~Erhan, I.~J. Goodfellow, and
  R.~Fergus.
\newblock Intriguing properties of neural networks.
\newblock In \emph{Proceedings of the 2nd International Conference on Learning
  Representations, {ICLR}}, 2014.

\bibitem[Urbano(2011)]{Urbano2011b}
J.~Urbano.
\newblock Information retrieval meta-evaluation: Challenges and opportunities
  in the music domain.
\newblock In \emph{Proceedings of the 12th Conference of the International
  Society for Music Information Retrieval}, pages 609--614, 2011.

\bibitem[Urbano and Flexer(2018)]{urbano2018statistical}
J.~Urbano and A.~Flexer.
\newblock Statistical analysis of results in music information retrieval: why
  and how (abstract).
\newblock In \emph{Proceedings of the International Society for Music
  Information Retrieval Conference}, pages xli--xlii, 2018.

\bibitem[Urbano et~al.(2012)Urbano, Downie, Mcfee, and
  Schedl]{urbano2012significant}
J.~Urbano, J.~S. Downie, B.~Mcfee, and M.~Schedl.
\newblock How significant is statistically significant? the case of audio music
  similarity and retrieval.
\newblock In \emph{Proceedings of the 13th Conference of the International
  Society for Music Information Retrieval}, pages 181--186, 2012.

\bibitem[Urbano et~al.(2013)Urbano, Schedl, and Serra]{urbano:etal:2013}
J.~Urbano, M.~Schedl, and X.~Serra.
\newblock Evaluation in music information retrieval.
\newblock \emph{Journal of Intelligent Information Systems}, 41\penalty0
  (3):\penalty0 345--369, 2013.

\bibitem[Urbano et~al.(2014)Urbano, Bogdanov, Boyer, G{\'o}mez~Guti{\'e}rrez,
  Serra, et~al.]{urbano2014effect}
J.~Urbano, D.~Bogdanov, H.~Boyer, E.~G{\'o}mez~Guti{\'e}rrez, X.~Serra, et~al.
\newblock What is the effect of audio quality on the robustness of mfccs and
  chroma features?
\newblock In \emph{Proceedings of the 15th Conference of the International
  Society for Music Information Retrieval}, pages 573--578, 2014.

\bibitem[Voorhees(2001)]{Voorhees2001a}
E.~M. Voorhees.
\newblock The philosophy of information retrieval evaluation.
\newblock In \emph{Proceedings of the Cross-Language Evaluation Forum}, 2001.

\bibitem[Wang(2003)]{Wang2003}
A.~Wang.
\newblock An industrial strength audio search algorithm.
\newblock In \emph{Proceedings of the 4th Conference of the International
  Society for Music Information Retrieval}, pages 7--13, Oct. 2003.

\bibitem[Wiggins(2009)]{wiggins:2009}
G.~A. Wiggins.
\newblock Semantic gap?? schemantic schmap!! methodological considerations in
  the scientific study of music.
\newblock In \emph{Proceedings of the 11th IEEE International Symposium on
  Multimedia}, pages 477--482. IEEE, 2009.

\end{thebibliography}

\end{document}